\documentclass[fleqn,usenatbib]{mnras}

\usepackage{newtxtext,newtxmath}

\usepackage[T1]{fontenc}
\usepackage{ae,aecompl}

\usepackage{graphicx}	
\usepackage{amsmath}	
\usepackage{amssymb}	
\usepackage{bm}
\usepackage{xcolor}

\title[Fundamental physics with blazar spectra]{Fundamental physics with blazar spectra: a critical appraisal}

\author[G. Galanti et al.]{
Giorgio Galanti,$^{1}$\thanks{E-mail: gam.galanti@gmail.com (GG)}
Fabrizio Tavecchio$^{1}$
and Marco Landoni$^{1}$
\\
$^{1}$INAF, Osservatorio Astronomico di Brera, Via E. Bianchi 46, I -- 23807 Merate, Italy
}

\date{Accepted XXX. Received YYY; in original form ZZZ}

\pubyear{2019}

\begin{document}
\label{firstpage}
\pagerange{\pageref{firstpage}--\pageref{lastpage}}
\maketitle

\begin{abstract}
Very-high-energy (VHE) BL Lac spectra extending above $10 \, \rm TeV$ provide a unique opportunity for testing physics beyond the standard model of elementary particle and alternative blazar emission models. We consider the hadron beam, the photon to axion-like particle (ALP) conversion, and the Lorentz invariance violation (LIV) scenarios by analyzing their consequences and induced modifications to BL Lac spectra. In particular, we consider how different processes can provide similar spectral features (e.g. hard tails) and we discuss the ways they can be disentangled. We use HEGRA data of a high state of Markarian 501 and the HESS spectrum of the extreme BL Lac (EHBL) 1ES 0229+200. In addition, we consider two hypothetical EHBLs similar to 1ES 0229+200 located at redshifts $z=0.3$ and $z=0.5$. We observe that both the hadron beam and the photon-ALP oscillations predict a hard tail extending to energies larger than those possible in the standard scenario. Photon-ALP interaction predicts a peak in the spectra of distant BL Lacs at about $20-30 \, \rm TeV$, while LIV produces a strong peak in all BL Lac spectra around $\sim 100 \, \rm TeV$. The peculiar feature of the photon-ALP conversion model is the production of oscillations in the spectral energy distribution, so that its detection/absence can be exploited to distinguish among the considered models. The above mentioned features coming from the three models may be detected by the upcoming Cherenkov Telescope Array (CTA). Thus, future observations of BL Lac spectra could eventually shed light about new physics and alternative blazar emission models, driving fundamental research towards a specific direction.
\end{abstract}

\begin{keywords}
astroparticle physics -- radiation mechanisms: non-thermal -- BL Lacertae objects: general -- galaxies: jets -- gamma-rays: galaxies.
\end{keywords}




\section{Introduction}

Radio-loud active galactic nuclei (AGNs) are extragalactic accreting supermassive black holes (SMBHs) characterized by the formation of two collimated relativistic jets that develop in opposite directions. It is commonly accepted that the basic structure of observed AGNs is similar: the different associated phenomenology is  mostly due to the different angle of view with which each source is observed at the Earth~\citep{urry95}. When merely by chance one of the AGN jets is almost pointing toward the Earth, radio-loud AGNs are called blazars. Their emission spans the entire electromagnetic spectrum, from radio waves up to the very-high-energy (VHE) gamma-ray band. Blazar spectra are characterized by two broad bumps, the first one peaking in the IR-UV band, whose origin is the synchrotron emission from relativistic electrons in the jet, while the second one reaches its maximum at gamma-ray energies. The origin of this high-energy bump is still debated. Leptonic models~\citep[e.g.][]{Maraschi92,Sikora94,ssc1} attribute it to inverse Compton scattering of synchroton radiation or external photons from the disc and/or clouds with the synchrotron emitting relativistic electrons. Hadronic models explain the photon high-energy emission as due to proton synchrotron emission or due to photomeson production (see e.g. ~\citealt{mannheim1,mannheim2,Muecke2003}).

Blazars are divided into two classes: flat spectrum radio quasars (FSRQs) and BL Lac objects (BL Lacs). FSRQs are powerful sources believed to be fed by an efficient accretion disc; moreover, they show optical emission lines produced by clouds of gas orbiting around the central SMBH and photoionized by UV photons from the disc. BL Lacs are less powerful and they do not display significant emission lines. These sources are believed to host a radiatively inefficient accretion flow unable to strongly ionize  the gas surrounding the SMBH (e.g.~\citealt{gmt09,righi19}). Moreover, as described by the so-called {\it blazar sequence}~\citep{fossati98,blazseq2}, BL Lac intrinsic gamma-ray spectra are generally harder than the FSRQs. In addition, in the latter case VHE photons are absorbed via the $\gamma\gamma \to e^+e^-$ process when interacting with soft background photons emitted by the broad-line region (BLR) and the dusty torus.

Deviations from the standard spectra predicted by leptonic and/or hadronic models, induced by conventional or new physics effects, are expected at VHE. In this fashion, the best obvious candidates are represented by BL Lacs: this is the reason why we concentrate on this blazar subclass in the following. The detection of photons coming from close BL Lacs ($z<0.1$) with energies above $20-30 \, \rm TeV$ and/or of photons from far BL Lacs ($z>0.3$) with energies well above $1 \, \rm TeV$ would be strong evidence that the current emission and/or propagation models of photons produced by BL Lacs are at least incomplete. The biggest challenge posed by extremely high energies and/or large distances is that  high-energy photons suffer severe absorption interacting with the optical-UV radiation of the extragalactic background light (EBL, see e.g.~\citealt{dgr2013,dwek13,gtpr}). Several scenarios are envisioned to interpret such a possible detection. Starting from conventional physics, in hadronic models and in the absence of strong extragalactic magnetic fields, protons may travel and produce an electromagnetic cascade which may result in a hardening of the observed spectra that can be detected at energies up to $20-30 \, \rm TeV$~\citep[e.g.][]{esseykusenko2010,esseyHB,MuraseEMC}. Another possibility is that physics beyond the Standard Model (SM) of elementary particle physics, with the introduction of axion-like particles (ALPs), may extend the detectable BL Lac spectra at similar energies~\citep{gtre} thanks to photon-ALP oscillations which increase the effective Universe transparency~\citep{grExt}. Alternatively, Lorentz invariance violation (LIV) would modify the photon dispersion relation and, as a consequence, would produce an enhancement of the photon flux for energies above $20 \, \rm TeV$~\citep{kifune,liv2}. In the following, we will cursorily sketch the main properties of the previous three models and then we will critically compare their predictions for BL Lac spectra by analyzing three sources: Markarian 501, 1ES 0229+200 and two hypothetical BL Lacs similar to the latter but placed at redshifts $z=0.3$ and $z=0.5$.  In particular, we study the effects of the three considered scenarios on the observed BL Lac spectra, discussing their phenomenological analogies and differences. A quite important point, overlooked in current discussions, is that the hadron beam and the photon-ALP oscillations produce a quite similar observational feature, i.e. a photon excess in observed BL Lac spectra at high energies. 
A similar degeneracy exists for LIV spectra, predicting a strong peak around $\sim 100 \, \rm TeV$, and photon-ALP oscillations in far-away sources. We will remark that the detection/absence of a spectral oscillatory behavior would discriminate among all scenarios, since that feature is  peculiar to the photon-ALP model. In addition, we show that, should the photon-ALP induced spectral oscillatory behavior be detected, the photon-ALP predictions may give also an indication about the blazar emission mechanism.

Although BL Lacs are the best candidates for testing modified models and new physics, even recent observations of FSRQs up to energies above $20 \, \rm GeV$ challenge current AGN models. In order to explain photon detection from FSRQs up to $\sim 300 \, \rm GeV$, standard AGN models must be modified by placing the emission region far from the center (see e.g.~\citealt{costamante18}) or it is possible to maintain classical AGN models by invoking photon-ALP conversion~\citep{trgb2012}. 

Our analysis of new-physics-induced BL Lac spectral modification in the VHE band is of great importance for the oncoming Cherenkov Telescope Array (CTA)\footnote{https://www.cta-observatory.org/} since it will be able to detect and precisely measure photons with energies up to above $100 \, \rm TeV$ and it may shed light on deviations from standard models of photon emission and/or propagation and/or even give proof of physics beyond the SM. Although CTA represents the most promising observatory for such studies, also current Imaging Atmospheric Che\-ren\-kov Telescopes (IACTs) H.E.S.S. (High Energy Stereoscopic System)\footnote{https://www.mpi-hd.mpg.de/hfm/HESS}, MAGIC (Major Atmospheric Gamma Imaging Cherenkov Telescopes)\footnote{https://magic.mpp.mpg.de/} and VERITAS (Very Energetic Radiation Imaging Telescope Array System)\footnote{http://veritas.sao.arizona.edu/} and other gamma-ray facilities like HAWC (High-Altitude Water Cherenkov Observatory)\footnote{http://www.hawc-observatory.org/}, GAMMA-400 (Gamma Astronomical Multifunctional Modular Apparatus)\footnote{gamma400.lebedev.ru/gamma$400_{\rm e}$.html}, LHAASO (Large High Altitude Air Shower Observatory)\footnote{http://english.ihep.cas.cn/ic/ip/LHAASO/}, TAIGA-HiSCORE (Hundred Square km Cosmic Origin Explorer)\footnote{www.desy.de/groups/astroparticle/score/en/} and HERD (High Energy Cosmic Radiation Detection,~\citealt{herd}) may make a detection.

The paper is organized as follows. In Sect. 2 we describe the hadron beam model, in Sect. 3 we sketch the main properties and consequences of the photon-ALP interaction, in Sect. 4 we illustrate LIV and its effects, while we critically compare the byproducts of the three scenarios for BL Lac spectra in Sect. 5 and we draw our conclusions in Sect. 6.

\section{Hadron beam scenario}
Standard hadronic models explain the second bump at gamma-ray energies as due to synchrotron emission by protons in the jet and/or photomeson production which produces gamma-ray photons due to neutral pion decay. A variant of the hadron model envisages that hadrons accelerated in the jet can escape in the form of a collimated beam of ultra-relativistic hadrons which, interacting with low-energy backgrounds, triggers electromagnetic cascades in extragalactic space~\citep[e.g.][]{esseykusenko2010,esseyHB,MuraseEMC}.


Cosmic ray-induced cascades may result in a hardening of the observed photon spectrum since the secondary photons produced suffer only a small path absorption due to the interaction with EBL photons. However, some important caveats must be taken into account. Primary cosmic rays are deflected by intense and structured magnetic fields. This is obviously the case for cluster and filament magnetic fields for cosmic rays with energies $E \lesssim 10^{19} \, \rm eV$~\citep{MuraseEMC}. Furthermore, the produced cascade may be deflected by the extragalactic magnetic field $B_{\rm ext}$ so knowledge of its strength and structure is important. For $B_{\rm ext} \gtrsim 10^{-15} \, \rm G$ -- which is however not so far from its lower limit~\citep{BextLIM,BextLIM2,BextLIM3,newBext} -- and a coherence length $\lambda_{\rm coh}={\cal O}(1 \, \rm Mpc)$ the cosmic ray-induced cascade is so deflected that the hadron beam scenario would be unable to efficiently produce a surplus of photons at TeV energies~\citep{tavBext}. As a high value of $B_{\rm ext}$ isotropizes the cosmic ray distribution, it would increase the cosmic ray luminosity budget needed for a sizable production of secondary photons. In addition, the cascade radiation appears as hardly compatible with rapidly varying sources since the cascade process produces a time spread~\citep{MuraseEMC}. As a consequence, this model can be applied to 1ES 0229+200 and to similar extreme BL Lacs \citep[e.g.][]{tavHBsim} but not to the highly variable Markarian 501.

Taking into account the above-mentioned remarks, the hadron beam scenario predicts a hardening of the observed spectra that can be detected at energies up to $20-30 \, \rm TeV$~\citep{tavHBsim}. The effect is more and more evident for more and more distant source: in particular, for closer sources ($z \lesssim 0.3$) we expect a hard tail above $\sim 10 \, \rm TeV$, while for sources at larger redshift the tail is predicted at lower energies because of the high attenuation (e.g.~\citealt{MuraseEMC}).

\section{Axion-like particles}
Axion-like particles (ALPs) are spin-zero, neutral, and very light pseudo-scalar bosons which are predicted by many extensions of the SM of particle physics such as String Theory (e.g.~\citealt{alp1,alp2}). APLs are a generalization of the axion (for a review see~\citealt{axion1,axion2,axion3}) -- the pseudo-Goldstone boson arising from the breakdown of the global Peccei-Quinn symmetry ${\rm U}(1)_{\rm PQ}$ which was proposed as a natural solution to the strong $CP$ problem. APLs differ from the original axion in two aspects. First, ALPs are supposed to interact primarily only with two photons: as a consequence, the additional Lagrangian~\footnote{We employ natural units in this paper.} describing the ALP field $a$ and ALP interactions with the SM reads 
\begin{equation}
\label{t1}
{\cal L}_{\rm ALP} = \frac{1}{2} \, \partial^{\mu} a \, \partial_{\mu} a - \, \frac{1}{2} \, m_a^2 \, a^2 + g_{a \gamma \gamma} \, a \, {\bm E} \cdot {\bm B}~,
\end{equation}
where ${\bm E}$ and ${\bm B}$ denote the electric and magnetic components of the electromagnetic tensor $F^{\mu \nu}$, while $m_a$ and $g_{a\gamma\gamma}$ are the ALP mass and the two-photon coupling of $a$, respectively. Furthermore, unlike the original axion, for ALPs $m_a$ and $g_{a \gamma \gamma}$ are totally unrelated quantities. Many bounds on $m_a$ and $g_{a \gamma \gamma}$ exist in the literature (for a thorough discussion see~\citealt{grExt}) but reasonable values appear to be $m_a = {\cal O}(10^{- 10} \, {\rm eV})$ and $g_{a \gamma \gamma} = {\cal O}(10^{- 11} \, {\rm GeV}^{- 1})$. In the presence of an external magnetic field, denoted by $\bm B$, photon-ALP oscillations may occur (in Eq.~(\ref{t1}) $\bm E$ represents the propagating photon field) since the propagation eigenstates differ from the interaction eigenstates being the mass matrix of the $\gamma - a$ system off-diagonal. Only the component ${\bm B}_T$ of ${\bm B}$ which is transverse to the photon momentum ${\bm k}$ couples to $a$ since ${\bm B}_T$ belongs to the plane containing $\bm E$ (see e.g.~\citealt{dgrx,grSM,grExt}). Therefore, every magnetized environment represents a candidate for photon-ALP oscillations. An improved and physically meaningful at all energies modeling of photon-ALP oscillations inside domain-like magnetic fields is discussed in~\citet{grSM}.

ALPs would have deep impact in astrophysics and especially in the VHE band whenever magnetic fields are intense and/or the path inside a magnetized medium is long (for an incomplete review see~\citealt{gRew}).  In particular, ALPs are produced in the magnetic field of the jet, so modifying BL Lac spectra (see \citet{trg}, also later in the text) and explaining FSRQ emission above $20 \, \rm GeV$ without placing the emission region far from the center and thus maintaining the validity of classical AGN models~\citep{trgb2012}. When a strong jet magnetic field is considered, ${\cal O}(1 \, \rm G)$, also the one-loop QED vacuum polarization must be accounted for~\citep{rs1988}; such magnetic field strengths are typical in hadronic models for blazar emission. Inside a galaxy cluster magnetic field, ALPs can produce irregularities in observed spectra which have, however, not been observed yet~\citep{fermi2016}. In extragalactic space photon-ALP oscillations can increase the Universe transparency to VHE photons by mitigating their attenuation via interaction with the photons of the EBL~\citep{dgrx,grExt}. Since in extragalactic space the magnetic field $B_{\rm ext}$ is weak, in the range $10^{-17} \, {\rm G} \lesssim B_{\rm ext} \lesssim 10^{-9} \, {\rm G}$ (e.g.~\citealt{BextLIM2}), also the photon dispersion on the cosmic microwave background (CMB) must be taken into account~\citep{raffelt2015,grExt}. Furthermore, photon-ALP oscillations have been employed tentatively to describe better the redshift dependence of AGN spectral indices~\citep{dgrx,horns12,troitsky14,bignami19} and to search for a diffuse flux of photons from ALP-to-photon back-conversation concomitant with neutrino production in extragalactic space~\citep{laha17}.

By combining photon-ALP conversion in the source, in extragalactic space and in the Milky Way we can obtain the BL Lac observed spectra modified by the existence of an ALP~\citep{gtre}. Hence, peculiar {\it features} arise: oscillations in the spectral energy distribution (SED) and photon excess above $20 \, \rm TeV$. In particular, for far away BL Lacs photon-ALP interaction produces an unexpected peak around $10-30 \, \rm TeV$, which is several orders of magnitude higher than conventional physics predictions. Thus, the analysis of BL Lac spectral features represents one of the best environments among the many studied for ALP physics to detect eventually such an ALP. A final remark is needed. Photon-ALP conversion inside the source depends only on the value of the jet magnetic field but it is independent of the particular state of the BL Lac: thus, the photon-ALP model can be applied equally well both to steady state and to flaring BL Lacs, contrary to the hadron beam model. As a result, we can apply ALP-induced modification to BL Lac spectra to all considered sources: Markarian 501, 1ES 0229+200 and the two BL Lacs similar to 1ES 0229+200 but located at redshifts $z=0.3$ and $z=0.5$.

\section{Lorentz invariance violation}
Many attempts to extend General Relativity to a quantum theory of gravity predict Lorentz invariance violation (LIV) beyond a very high energy scale $E_{\rm LIV}$. Depending on the model parameters, LIV turns out to possess a very rich phenomenology modifying standard physics interactions and also allowing for otherwise forbidden processes, such as: photon decay, vacuum Cherenkov effect, shifting of existing threshold reactions, photon splitting~\citep{liberati13}. Many sectors of the SM are affected by LIV-induced modified dispersion relations: recently, LIV impact on neutrino oscillations has been studied in~\citet{LIVneutrino} by using the approach originally introduced by~\citet{coleman}.

Here, we are interested in LIV consequences for photon propagation that translate into a photon modified dispersion relation that contains additional terms in the form of $E^{n+2}/E^n_{\rm LIV}$ where $E$ represents the photon energy (see e.g~\citealt{kifune}). Limiting to the case $n=1$, the modified dispersion relation for photons reads
\begin{equation}
\label{t2}
E^2-p^2=-\frac{E^3}{E_{\rm LIV}}~,
\end{equation}
where $p$ is the photon momentum. The term in the right hand side of Eq.~(\ref{t2}) $m_{\gamma, {\rm eff}}^2 \equiv -E^3/E_{\rm LIV}$ formally represents a photon effective mass term which induces a modification in the threshold of the process $\gamma\gamma\to e^+e^-$. Once the parameter $E_{\rm LIV}$ -- which is of the order of the Planck energy -- is fixed, the effect of $m_{\gamma, {\rm eff}}^2$ is more and more dramatic as the energy grows as can be inferred from  Eq.~(\ref{t2}). Although the calculation of the optical depth associated with the process $\gamma\gamma\to e^+e^-$ and accounting for the LIV effect is not trivial and possesses some uncertainties coming from different possible assumptions~\citep{fairbainLIV,jacobLIV}, the final result is not deeply affected by such uncertainties. Therefore, it is possible to state that the modifications induced by the LIV start to be sizable for $E \gtrsim {\cal O}(10 \, \rm TeV)$: the resulting effective transparency turns out to be larger than the standard one~\citep{tavLIV}. Since such modifications are expected for $E \gtrsim {\cal O}(10 \, \rm TeV)$ the best astrophysical sources to test such a scenario appear to be high-frequency peaked BL Lacs (HBLs) and extreme BL Lacs (EHBLs) since their observed spectra extend to few TeV~\citep{steckerLIV,tavLIV}. As in the case of photon-ALP oscillations, LIV effects on photon transparency are totally independent of the particular state or characteristics of the BL Lacs. Consequently, the LIV model can be applied both to steady state and to flaring BL Lacs, unlike the hadron beam model. Thus, we analyze LIV-induced modifications to the high-energy spectra of Markarian 501, 1ES 0229+200 and two BL Lacs similar to 1ES 0229+200 but placed at redshifts $z=0.3$ and $z=0.5$.

\section{Analysis of modified BL Lac spectra}
In this section, we describe the effects of the hadron beam, the photon-ALP oscillations and LIV on BL Lac spectra and we compare the three scenarios, pointing out phenomenological analogies and differences in order to clarify how present and future observational data might detect a hint for new physics coming from one of these three models and how to discriminate among them. Therefore, we analyze hereafter the best candidates for these purposes since they present a very powerful spectrum up to above $10 \, \rm TeV$: Markarian 501, 1ES 0229+200 and two BL Lacs similar to 1ES 0229+200 but placed at redshifts $z=0.3$ and $z=0.5$. As benchmark values we take $m_a=10^{-10} \, \rm eV$ and $g_{a\gamma\gamma}=10^{-11} \, \rm GeV^{-1}$ for the photon-ALP model and $E_{\rm LIV}=10^{20} \, \rm GeV$ for the LIV scenario \citep{LIVlimit}. As far as the EBL is concerned, we employ the model by~\citet{EBLfranc}.



\subsection{Markarian 501}
\begin{figure}   
\begin{center}
\includegraphics[width=.5\textwidth]{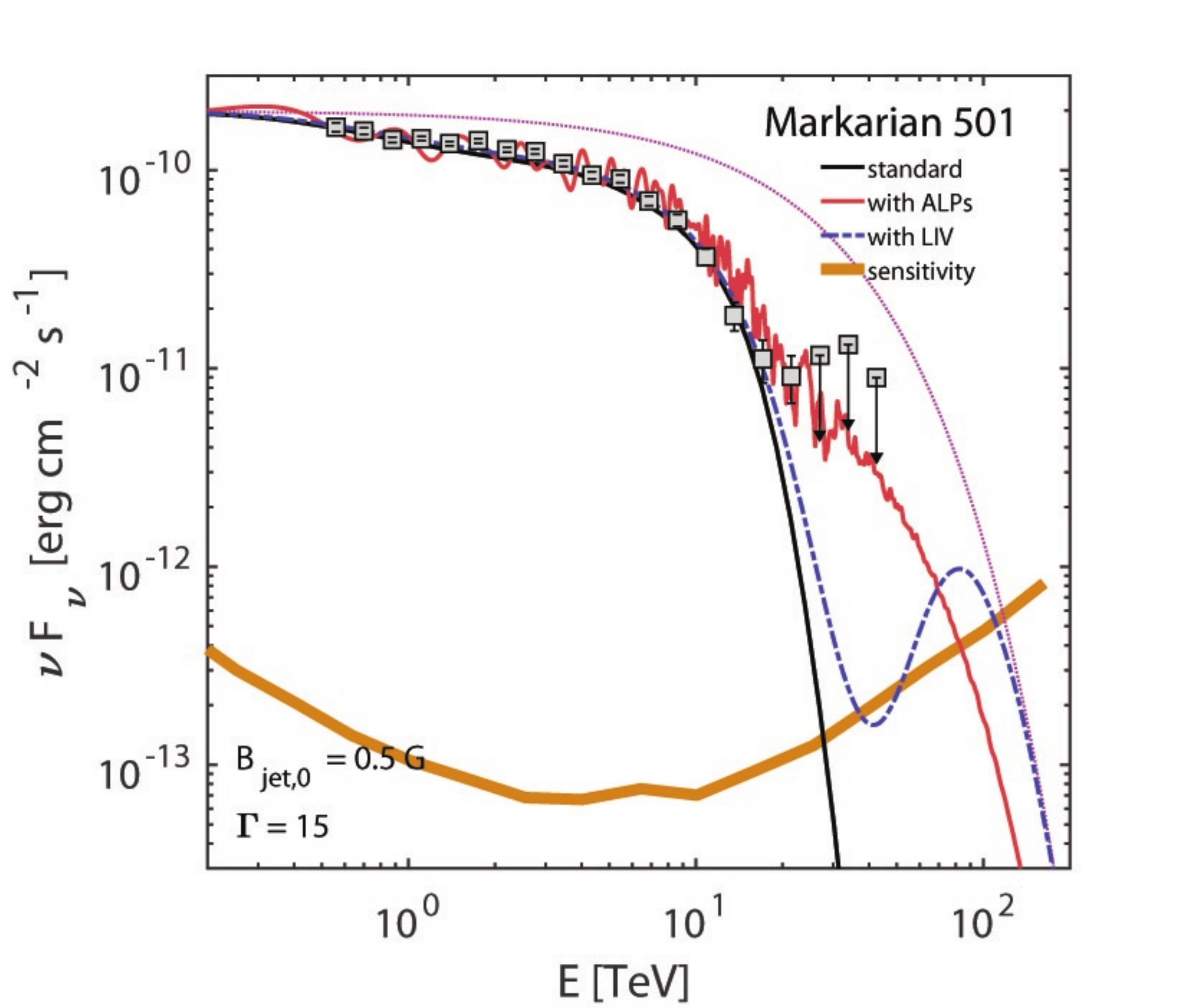}
\end{center}
\caption{\label{Mrk501} Behaviour of the observed spectrum of Markarian 501 versus the observed energy $E$. The solid black line corresponds to conventional physics, the solid red line to the scenario of the photon-ALP oscillations and the dot-dashed blue line to the LIV model. The dotted purple line is the intrinsic exponentially truncated power law spectrum, and the solid orange line represents the CTA sensitivity for the south site and 50 h of observation. We take $B_{\rm jet,0}=0.5 \, \rm G$ and $\Gamma=15$. See the text for more details. The grey squares are the observational data detected by HEGRA~\citep{hegra}.}
\end{figure}

Markarian 501 is a nearby and luminous HBL  at a redshift $z=0.034$. Although its quiescent states have been studied as a possible environment where to discover signals of new physics~\citep{fairbainLIV}, active states of Markarian 501 are certainly the best `laboratory' for such an issue~\citep{tavLIV}. In this fashion, the observational data points from HEGRA~\citep{hegra} represent one of the best examples present in the literature which we use in the following: its high state reaches energies up to above $10 \, \rm TeV$ with a hard spectrum (see also the recent high state detected by HESS, \citealt{abdalla}). As already mentioned above, the hadron beam scenario cannot be applied in this case because of the variability of the source: thus, we concentrate here on photon-ALP interaction and LIV effects of the spectrum of Markarian 501, as reported in Fig.~\ref{Mrk501}. We take a jet magnetic field with a toroidal geometry that assumes the value $B_{\rm jet,0}=0.5 \, \rm G$ in the emission region and we assume a bulk Lorentz factor $\Gamma =15$ (see also \citealt{trg}). We envisage that the emission mechanism for this source is leptonic and of the Synchrotron Self-Compton (SSC) type. 

For this source we model the intrinsic spectrum with an exponentially truncated power law with energy index $\alpha_1=1$ and cutoff energy $E_{\rm cut}=20 \, \rm TeV$.
Fig.~\ref{Mrk501} shows that, as expected from~\citet{gtre}, photon-ALP oscillations induce an energy-dependent behaviour of the spectrum and a photon excess for energies above $\sim 20 \, \rm TeV$ as compared to conventional physics expectations. In agreement with~\citet{tavLIV}, LIV causes an hardening of the observed spectrum for energies above $\sim 20 \, \rm TeV$ in a similar way as photon-ALP interactions but LIV predicts also a possible minimum in the spectrum around $\sim 40 \, \rm TeV$ and a subsequent peak around $\sim 100 \, \rm TeV$ which are not present in the photon-ALP scenario.

The detection of a photon excess alone above $\sim 20 \, \rm TeV$ as compared to expectations from conventional physics would represent evidence for new physics in terms of the existence of an ALP or of the breakdown of Lorentz invariance: such an eventual detection, however, could not discriminate between the two models. Nevertheless, while a minimum followed by a peak in the spectrum above $\sim 20 \, \rm TeV$ cannot be viewed as a clear preference for the LIV scenario, the observation of energy-dependent oscillations in the spectrum would represent a clear smoking gun for the photon-ALP interaction scenario since the latter is the only one which predicts such oscillations in the spectrum. All the above-mentioned features driven by the photon-ALP and the LIV models appear to be detectable by the CTA with an observational exposure of $50 \, \rm h$~\citep{CTAsens}.


\subsection{1ES 0229+200}

\begin{figure}   
\begin{center}
\includegraphics[width=.5\textwidth]{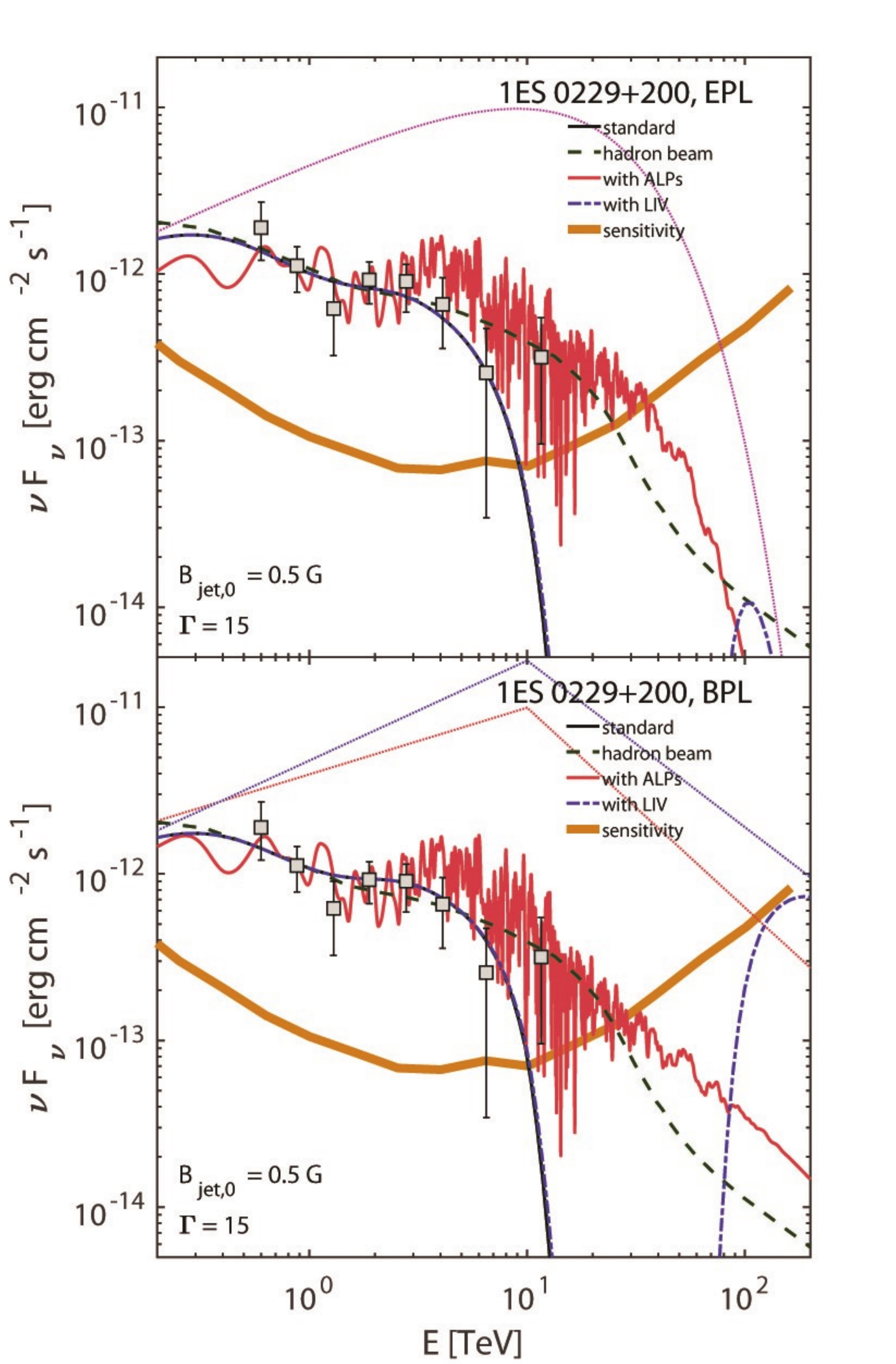}
\end{center}
\caption{\label{0229} Same as Fig.~\ref{Mrk501} but for 1ES 0229+200. In addition, the dashed green line corresponds to the hadron beam scenario. In the upper panel we consider an intrinsic exponentially truncated power law (EPL) spectrum (the same for the photon-ALP interaction scenario and the LIV one corresponding to the dotted purple line), while in the lower panel we take an intrinsic broken power law (BPL) spectrum (we take two different models for the photon-ALP interaction scenario corresponding to the dotted red line and the LIV one corresponding to the dotted blue line). We take $B_{\rm jet,0}=0.5 \, \rm G$ and $\Gamma=15$. See the text for more details. The grey squares are the observational data from HESS~\citep{hess0229}.}
\end{figure}
\begin{figure}   
\begin{center}
\includegraphics[width=.5\textwidth]{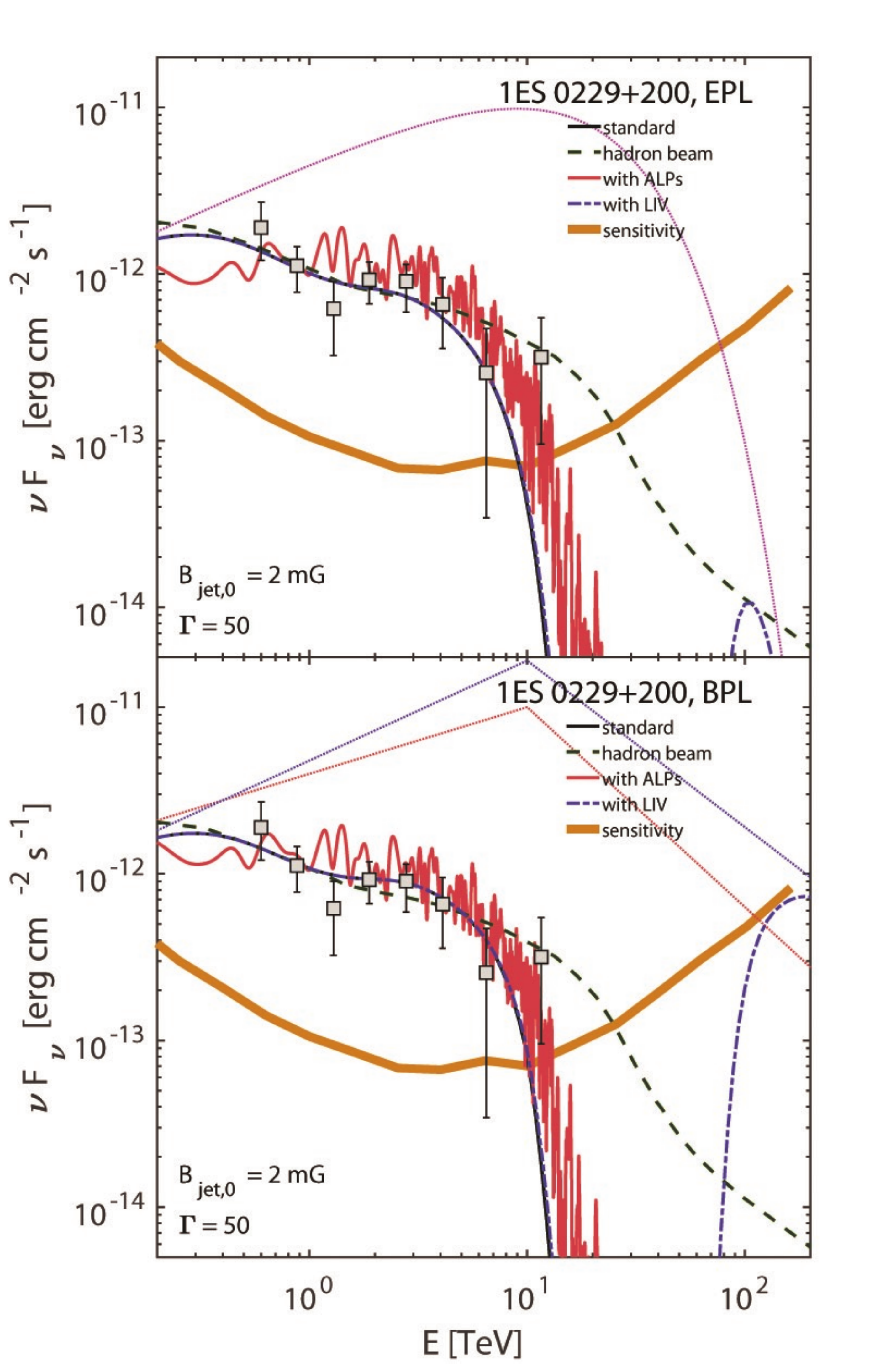}
\end{center}
\caption{\label{0229L} Same as Fig.~\ref{0229} but we take $B_{\rm jet,0}=2 \, \rm mG$ and $\Gamma=50$. See the text for more details.}
\end{figure}

1ES 0229+200 represents the prototype of EHBLs and possesses a hard TeV observed spectrum extending above $10 \, \rm TeV$~\citep{hess0229} in spite of the fact that it is located at a redshift $z=0.1396$, where EBL absorption in the TeV energy range is huge (optical depth $\tau \gtrsim 2$, e.g.~\citealt{EBLfranc}). EHBLs have peculiar and not yet completely understood features which differentiate them from other BL Lacs. As far as the VHE spectrum is concerned, its extreme hardness challenges standard one-zone leptonic models (see e.g.~\citealt{tavExtreme09}) and its weak variability is untypical for the BL Lac population (see e.g.~\citealt{variabExtreme}).

In Fig.~\ref{0229} we show the spectrum of 1ES 0229+200 modified by the inclusion of the hadron beam, photon-ALP and LIV effects, when we consider a jet toroidal magnetic field with a value $B_{\rm jet,0}=0.5 \, \rm G$ in the emission region and a bulk Lorentz factor $\Gamma =15$. These values are found when the emission is modeled with the hadronic scenario of the proton-synchrotron type (see e.g. \citealt{Cerruti15}) -- for simplicity we call this scenario `high magnetic field case'. In Fig.~\ref{0229L} we consider the same jet magnetic field behaviour but we take $B_{\rm jet,0}=2 \, \rm mG$ and $\Gamma =50$ by assuming a leptonic SSC emission mechanism (see e.g. \citealt{Costamante18b}) -- for simplicity we call this scenario `low magnetic field case'.

In the upper panel of Fig.~\ref{0229} we model the intrinsic spectrum with an exponentially truncated power law with energy index $\alpha_1=0.4$ and cutoff energy $E_{\rm cut}=15 \, \rm TeV$. In the lower panel we use a broken power law with different parameters for the photon-ALP conversion scenario and the LIV model: we have energy index $\alpha_1^{\rm ALP}=0.6$ and $\alpha_1^{\rm LIV}=0.4$, high-energy index $\alpha_2^{\rm ALP} =2.2$ and $\alpha_2^{\rm LIV} = 2$ and break energy $E_b^{\rm ALP} = E_b^{\rm LIV} = 10 \, \rm TeV$. From Fig.~\ref{0229} we observe that the hadron beam scenario produces a hard tail above $\sim 10 \, \rm TeV$ extending monotonically up to above $100 \, \rm TeV$ as already reported in~\citet{tavHBsim}. In agreement with~\citet{gtre} photon-ALP oscillations produce an energy-dependent behaviour of the spectrum and a sizable photon excess above $\sim 10 \, \rm TeV$ both when the intrinsic spectrum is an exponentially truncated power law and even more when it is a broken power law. As expected from~\citet{tavLIV}, LIV-induced modifications on the spectrum are negligible around $\sim 10 \, \rm TeV$ but they predict a peak around $\sim 100 \, \rm TeV$ that is hardly observable in the case of an intrinsic exponentially truncated spectrum but possibly detectable for an intrinsic broken power law one by considering the predicted sensitivity of the CTA~\citep{CTAsens}. The other features induced by the three models appear to be detectable with an observational exposure of $50 \, \rm h$.

In Fig.~\ref{0229L} we take the same intrinsic spectra as in Fig.~\ref{0229}. Everything we have observed in Fig.~\ref{0229} extends to Fig.~\ref{0229L} concerning the hadron beam and the LIV scenario. We note instead a modification of the spectrum induced by the photon-ALP oscillations: in the `low magnetic field case' the conversion inside the source is not very efficient and the hard tail in the observed spectrum is greatly reduced. In this case, the most evident imprinting of the photon-ALP interaction is the still remaining peculiar energy-dependent behaviour of the spectrum. The photon-ALP model is almost identical in the cases of both exponentially truncated (upper panel of Fig.~\ref{0229L}) and broken power law (lower panel of Fig.~\ref{0229L}) intrinsic spectrum.

The observation alone of a harder spectrum above $\sim 10 \, \rm TeV$ as compared to conventional physics predictions could be produced either by the hadron beam or by the photon-ALP oscillation model in the `high magnetic field case'. The smoking gun to distinguish between the two scenarios is again the eventual detection of energy-dependent oscillations in the observed spectrum since this feature is driven only by the photon-ALP oscillation scenario. Instead, only LIV can reproduce an eventual detection around $\sim 100 \, \rm TeV$.


\subsection{Extreme BL Lacs at $z=0.3$ and $z=0.5$}

Due to the observation of BL Lacs up to above a redshift $z \ge 0.5$ we speculate on the existence of EHBLs at redshifts $z=0.3$ and $z=0.5$. We assume characteristics similar to the prototype of EHBLs, 1ES 0229+200, but more extreme with respect to the cutoff or break energy. In particular, we use $E_{\rm cut}$ a factor 2 larger that partially compensates for the enhanced EBL absorption.

\begin{figure}   
\begin{center}
\includegraphics[width=.5\textwidth]{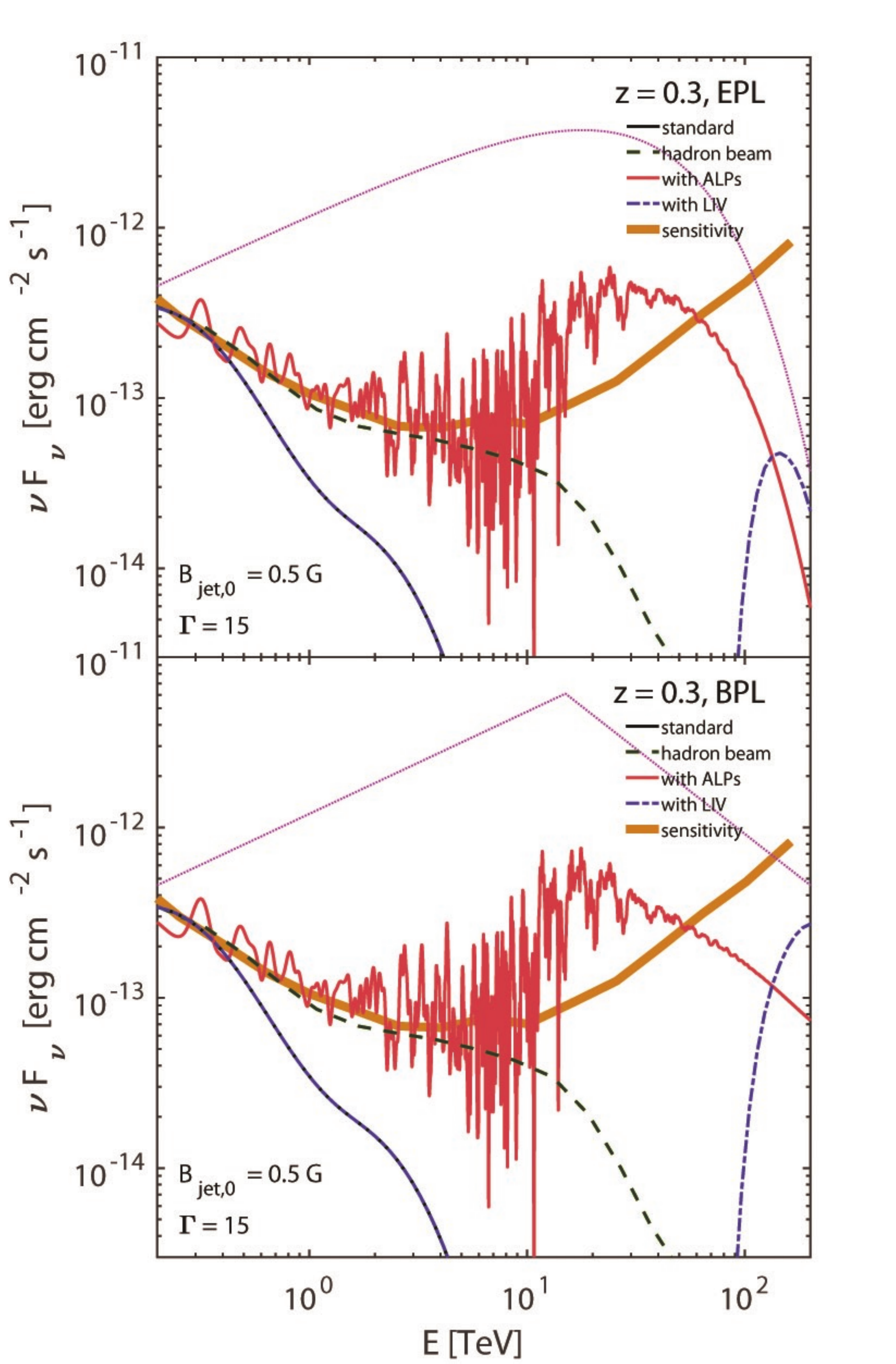}
\end{center}
\caption{\label{z03} Same as Fig.~\ref{0229} but for a BL Lac at $z = 0.3$. In this case for both the photon-ALP interaction and the LIV scenario we take the same parameters concerning the intrinsic exponentially truncated power law (upper panel) and the broken power law (lower panel) spectrum. We take $B_{\rm jet,0}=0.5 \, \rm G$ and $\Gamma=15$. See the text for more details.}
\end{figure}

\begin{figure}   
\begin{center}
\includegraphics[width=.5\textwidth]{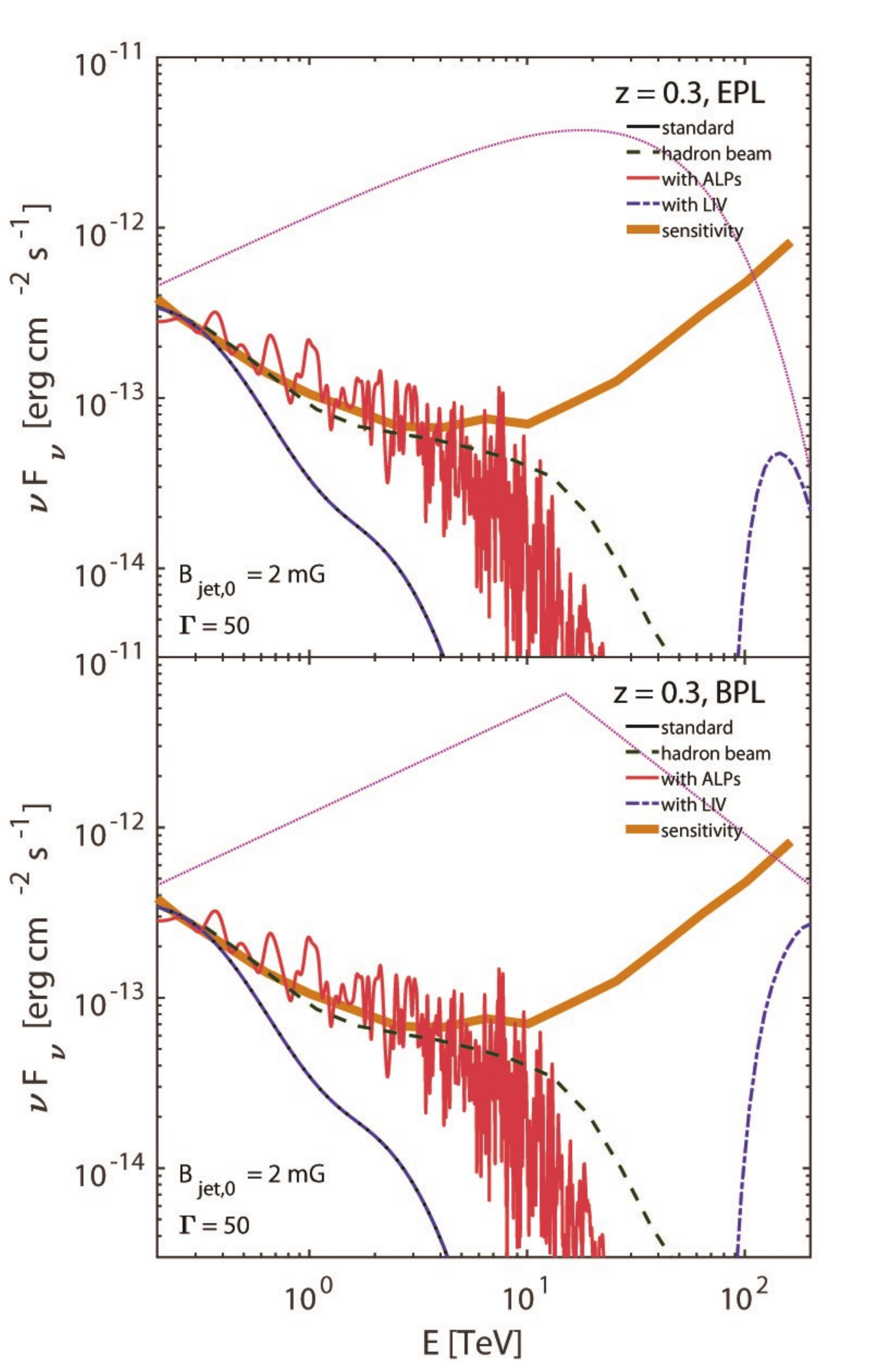}
\end{center}
\caption{\label{z03L} Same as Fig.~\ref{z03} but we take $B_{\rm jet,0}=2 \, \rm mG$ and $\Gamma=50$. See the text for more details.}
\end{figure}

In Fig.~\ref{z03} and Fig.~\ref{z03L} we report the spectrum of the considered BL Lac at reshift $z=0.3$ when the hadron beam, photon-ALP interaction and LIV effects are taken into account in the case of both the `high magnetic field case' (Fig.~\ref{z03}) and `low magnetic field case' (Fig.~\ref{z03L}) with the same parameters of 1ES 0229+200.

In the upper panel of Fig.~\ref{z03} for the intrinsic spectrum we use an exponentially truncated power law with energy index $\alpha_1=0.4$ and cutoff energy $E_{\rm cut}=30 \, \rm TeV$. In the lower panel we employ a broken power law with energy index $\alpha_1=0.4$, high-energy index $\alpha_2 = 2$ and break energy $E_b = 15 \, \rm TeV$. What we observed for 1ES 0229+200 still stands here only with little modifications. As compared to conventional physics the hadron beam model produces a hard tail above a few TeV. A similar photon excess is predicted by the photon-ALP interactions which as usual generates energy-dependent oscillations in the observed spectrum. In addition, the photon-ALP model gives rise to a peak in the observable spectra at energies around $20-30 \, \rm TeV$ in the case of both intrinsic exponentially truncated power law and broken power law spectrum. LIV modifications are confined to energies above $\sim 100 \, \rm TeV$, where a strong peak in the observed spectrum in expected. For lower energies LIV impact on the observable spectrum is negligible.

The assumed intrinsic spectra in Fig.~\ref{z03L} are the same of Fig.~\ref{z03}. As for 1ES 0229+200, the hadron beam and the LIV scenario effects in Fig.~\ref{z03L} are the same as in Fig.~\ref{z03}. Concerning the photon-ALP interaction, the `low magnetic field case' of Fig.~\ref{z03L} shows that the conversion inside the source is low thus not allowing for the production of a peak in the observed spectrum around $20-30 \, \rm TeV$ as predicted instead in the `high magnetic field case' of Fig.~\ref{z03}. In the `low magnetic field case' of Fig.~\ref{z03L} the hard tail up to $\sim 10 \, \rm TeV$ still remains.
Again, the signature of the photon-ALP interaction is the persisting peculiar energy-dependent behaviour of the spectrum. The different choice of the intrinsic spectrum with an exponentially truncated (upper panel of Fig.~\ref{z03L}) and broken power law (lower panel of Fig.~\ref{z03L}) does not significantly affect the predictions of the photon-ALP model.

We observe in Fig.~\ref{z03} that the hadron beam and the photon-ALP interactions produce a similar hardening in the observable spectrum up to $\sim 10 \, \rm TeV$ and once more the detection/absence of energy-dependent oscillations in the spectrum would represent the possibility to distinguish between the two models. For the considered BL Lac at redshift $z=0.3$ the hard tail up to $\sim 10 \, \rm TeV$ predicted both by the hadron beam and by the photon-ALP interaction is of the order of the CTA sensitivity and thus believed as observable. In addition, we expect that the peak predicted by the photon-ALP scenario around $20-30 \, \rm TeV$ in the `high magnetic field case' could be detected in the near future by the CTA~\citep{CTAsens}. Since other considered scenarios do not predict a peak at these energies, an eventual detection of photons at $20-30 \, \rm TeV$ would represent a smoking gun for the existence of an ALP. Although it appears less likely, a detection at energies above $100 \, \rm TeV$ would be a strong indication of LIV effects (with an intrinsic broken power law spectrum) since other models are unable to reach efficiently such energies (only partially this may be the case for photon-ALP interaction with an intrinsic broken power law spectrum).


\begin{figure}   
\begin{center}
\includegraphics[width=.5\textwidth]{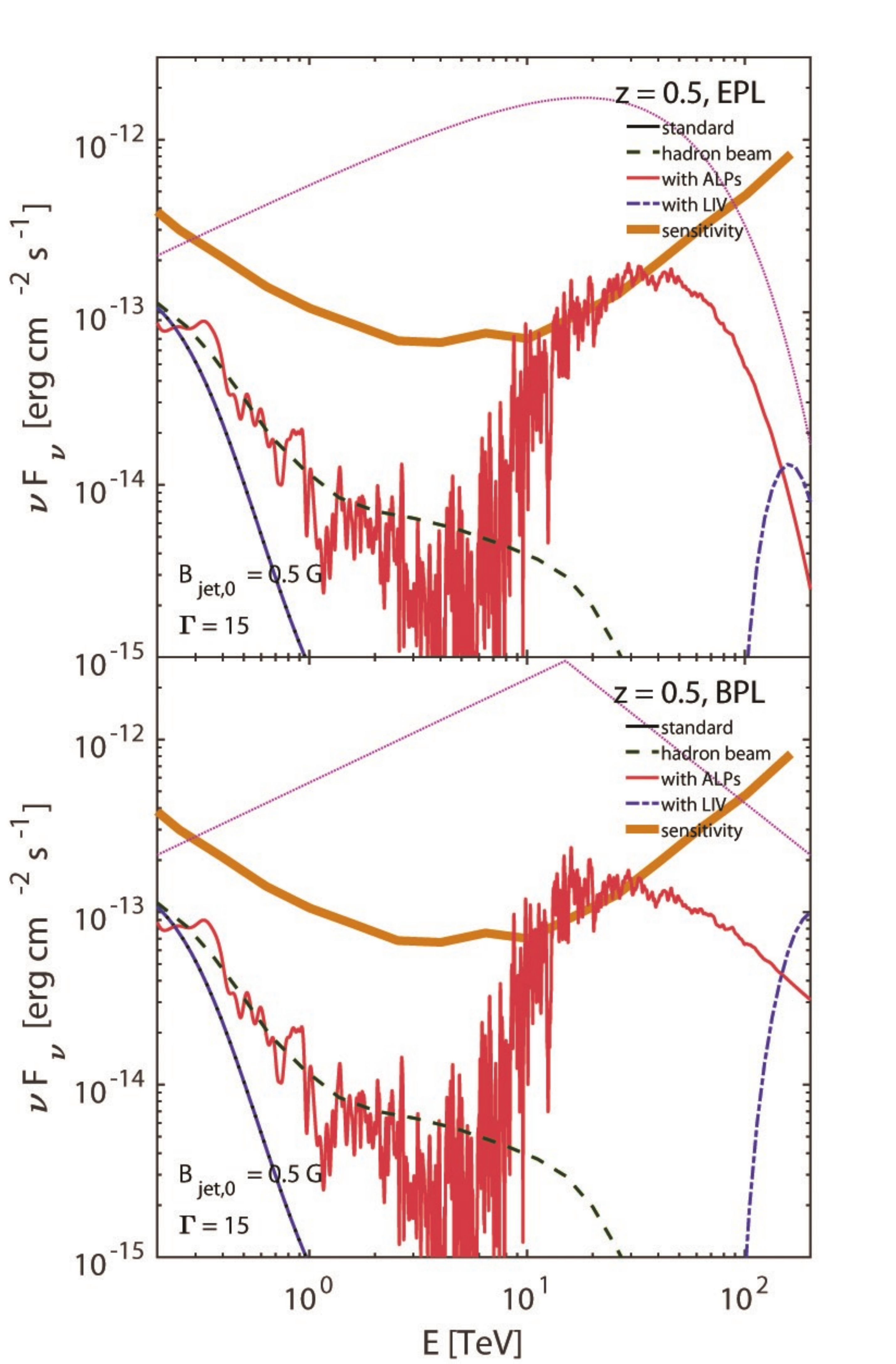}
\end{center}
\caption{\label{z05} Same as Fig.~\ref{z03} but for a BL Lac at $z = 0.5$. See the text for more details.}
\end{figure}

\begin{figure}   
\begin{center}
\includegraphics[width=.5\textwidth]{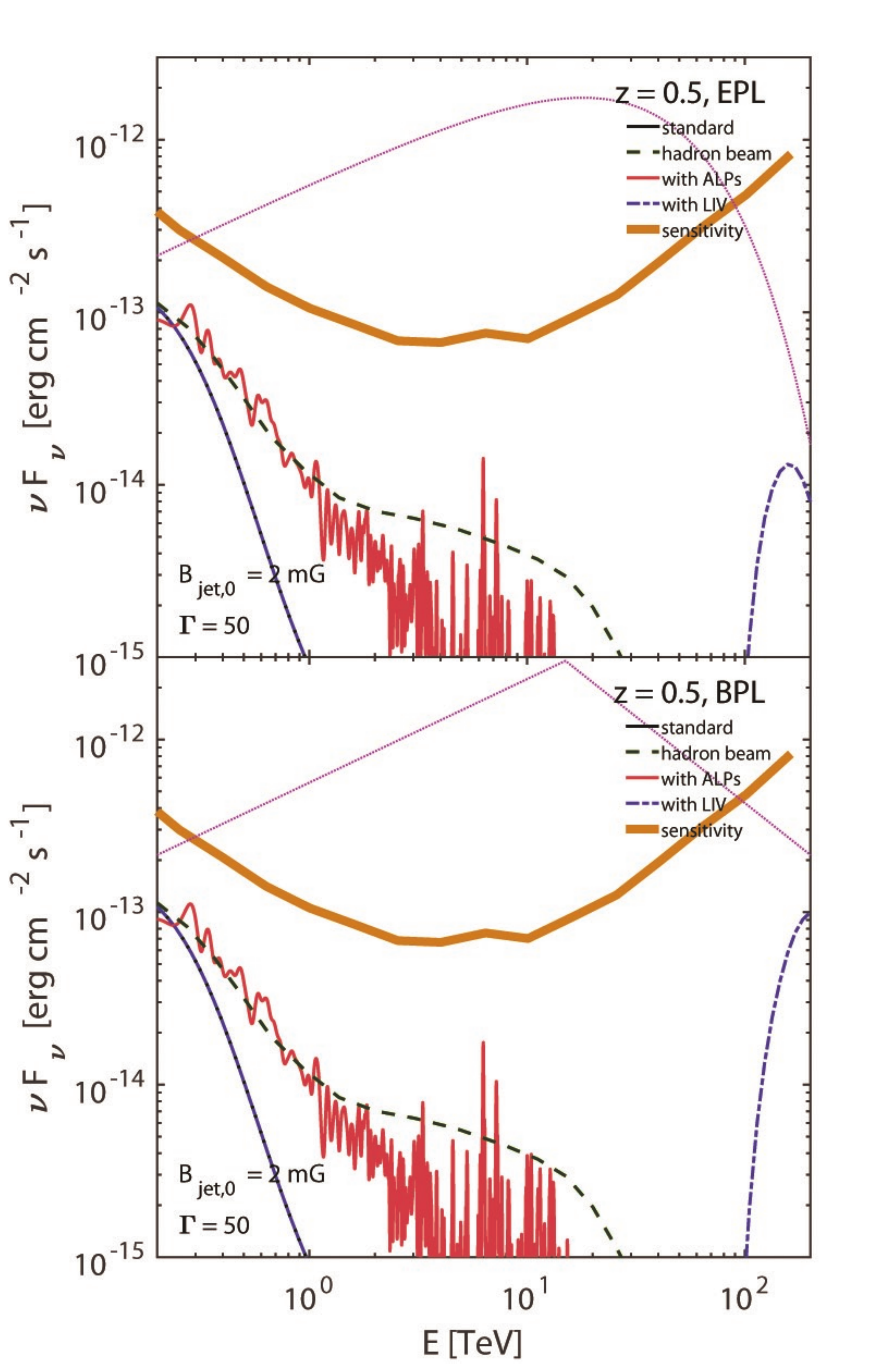}
\end{center}
\caption{\label{z05L} Same as Fig.~\ref{z03L} but for a BL Lac at $z = 0.5$. See the text for more details.}
\end{figure}

Everything we have stated for the $z=0.3$ case translate to the case of a BL Lac at redshift $z=0.5$, whose new-physics-induced modified spectra are reported in Fig.~\ref{z05} for the `high magnetic field case' and in Fig.~\ref{z05L} for the `low magnetic field case'. All parameters here are the same chosen for the case $z=0.3$. In any case, for such a far source the real detectability of these features remains problematic even for energies below $\sim 1 \, \rm TeV$. Instead, the peak predicted by the photon-ALP scenario around $20-30 \, \rm TeV$ in the `high magnetic field case' is expected to be observable by the CTA~\citep{CTAsens} -- which would unequivocally be associated with the existence of an ALP since other processes are unable to produce a similar peak. Instead, the detection of photons at energies above $100 \, \rm TeV$ would represent a rather robust hint at LIV effects (with an intrinsic broken power law spectrum) since photon-ALP interaction with an intrinsic broken power law spectrum might only partially produce consequences at such energies and the hadron beam scenario totally fails in this respect. However, such a detection -- already challenging in the case $z=0.3$ -- is considered as prohibitive in the case $z=0.5$ even for the CTA capabilities.



\section{Conclusions}
In this paper, we have analyzed the effects of the hadron beam scenario, photon-ALP oscillations and LIV on BL Lac spectra. We have concentrated on an archival high state of Markarian 501 and on 1ES 0229+200 since they represent the best candidates for a search for new physics signatures. In addition, we have considered two hypothetical EHBLs at redshifts $z=0.3$ and $z=0.5$. 

The hadron beam scenario -- which is not applicable to the high state of Markarian 501 because of the source high variability -- gives rise to a hard tail with respect to conventional physics for observed BL Lac spectra. A similar photon excess is predicted in the `high magnetic field case' by the photon-ALP interaction model but the two scenarios can be distinguished by searching for energy-dependent oscillations in the observed spectrum since they may be induced only by photon-ALP interaction. LIV gives rise to a peak in the spectra at energies around $\sim 100 \, \rm TeV$ for both close and far-away sources while a strong peak is predicted in the `high magnetic field case' by photon-ALP oscillations for farther sources only and at slightly lower energies ($\sim 20-30 \, \rm TeV$). In any case, the concomitant detection of energy oscillations in the spectrum would be the smoking gun for the photon-ALP interactions. As is evident from all figures, the amplitude of the photon-ALP induced energy oscillations in the spectrum becomes bigger and bigger as the redshift of the source grows: the reason is that the photon-ALP beam is sensitive to all crossed magnetized media. Thus, a more distant source crosses many different magnetic field structures and this fact amplifies the variation of the oscillation amplitude since the photon-ALP system keeps memory of all crossed magnetized media.

As a side note, we remark that the ALP identification through the detection of energy-dependent oscillations in the observed spectrum may represent, at least in principle, a unique opportunity to infer also information about the emission mechanism of BL Lacs. Since the efficiency of the photon-ALP conversion is strictly related to the intensity of the magnetic field crossed by the photon-ALP beam, a very hard tail above $\sim 20 \, \rm TeV$ observed for 1ES 0229+200 would imply a very high value for the magnetic field in the jet, suggesting a hadronic emission mechanism. Instead, the absence of a hard tail above $\sim 20 \, \rm TeV$ would be related to a lower value for the magnetic field in the jet, suggesting a leptonic SSC emission mechanism. A similar conclusion may be achieved about the hypothetical sources at redshifts $z=0.3$ and $z=0.5$: here, the distinction is represented by the presence/absence of a peak in the observed spectrum at $\sim 20-30 \, \rm TeV$. The eventual observation of the peak -- associated with a high value of the magnetic field in the jet -- would be an indication for a hadronic emission mechanism, while its absence would imply a preference for a leptonic SSC emission mechanism compatible with lower values of the jet magnetic field.

As far as the real detectability of such features is concerned, we stress that, although Markarian 501 and 1ES 0229+200 are northern sources, we show for all BL Lacs the CTA south sensitivity (where such sources are observable with a large zenith angle). The reason for this choice is that, since we are interested in non-SM effects in the TeV and multi-TeV range, as a matter of fact, the best telescopes suitable for these observations are the SSTs (Small-Sized Telescopes), foreseen only in the south site. In addition, the instrument sensitivity turns out to be greatly increased in and above the TeV energy range (at the expense of the increase of the low energy threshold) at large ($\gtrsim 60^{\circ}$) zenith angles~\citep{iatcHighZenith}. We base our considerations on the predicted CTA sensitivity~\citep{CTAsens}: both the photon excess and the oscillatory behavior of the spectrum may be observed for closer sources, while for the further ones only a very long exposure ($> 50 \, \rm h$) for steady state sources may allow for a detection. In particular, concerning the detectability by the CTA of ALP-induced spectral energy oscillations, we expect that they may be discriminated for energies below $\sim 5 \, \rm TeV$ especially for closer sources since the number of oscillations per energy decade is low enough (see Figs.~\ref{Mrk501}-\ref{0229L}) as compared the CTA energy resolution. In any case, dedicated simulations are required in order to test the real detectability of this feature. The peak at $\sim 20-30 \, \rm TeV$ induced by the photon-ALP interaction model in the `high magnetic field case' appears to be detectable even if present for far-away sources while the LIV-induced peak at around $\sim 100 \, \rm TeV$ appears to be observable for close sources and hard intrinsic spectra while for the far-away ones a long exposure time ($> 50 \, \rm h$) for steady state sources is necessary. We want to stress that a long time of exposure for closer steady state sources even for energies below $\sim 10 \, \rm TeV$ would be crucial to decrease flux error bars in order to detect eventually photon-ALP induced energy oscillations in the spectra. We expect that all these observations may be performed by the upcoming CTA.

Finally, in spite of the fact that CTA is expected to be the most promising facility for observing the features considered in this paper, also present IACTs H.E.S.S., MAGIC and VERITAS and other gamma-ray observatories like HAWC, GAMMA-400, LHAASO, TAIGA-HiSCORE and HERD may give an indication of new physics.


\section*{Acknowledgments}
We thank the referee for useful comments. GG and FT acknowledge contribution from the grant INAF CTA--SKA, ``Probing particle acceleration and $\gamma$-ray propagation with CTA and its precursors" and the INAF Main Stream project ``High-energy extragalactic astrophysics: toward the Cherenkov Telescope Array''.

This paper has gone through internal review by the CTA Consortium.


\label{lastpage}

\end{document}